\documentclass[10pt,journal]{IEEEtran}
\usepackage[table]{xcolor}
\usepackage{cite}      
\usepackage{graphicx} 
\usepackage{amsmath}
\usepackage{fourier}
\usepackage[switch]{lineno}

\usepackage{psfrag}    
\usepackage{subfigure}
\usepackage{hyperref}
\usepackage{cleveref}
\usepackage{placeins}
\usepackage{multirow}
\usepackage{titlesec}
\titlespacing{\section}{0pc}{1pc}{0pc}
\titlespacing{\subsection}{0pc}{1pc}{0pc}
\usepackage{multicol}
\usepackage{lipsum}
\usepackage{url}
\hypersetup{
    colorlinks = true,
    linkcolor = black,
    citecolor= black,
    urlcolor =blue
  }
\usepackage{stfloats}  
\usepackage{amsmath}   
\usepackage{titlesec}
\titlespacing\section{0pt}{12pt plus 4pt minus 2pt}{0pt plus 2pt minus 2pt}
\titlespacing\subsection{0pt}{12pt plus 4pt minus 2pt}{0pt plus 2pt minus 2pt}
\titlespacing\subsubsection{0pt}{12pt plus 4pt minus 2pt}{0pt plus 2pt minus 2pt}

\usepackage{array}
\usepackage[rawfloats=true]{floatrow} 
\restylefloat{figure} 
\usepackage[font=small,skip=0pt]{caption}
\newcolumntype{M}[1]{>{\centering\arraybackslash}m{#1}}
\begin{document}
\title{Neural Style Transfer Enhanced Training Support For Human Activity Recognition}
\author{
    \IEEEauthorblockN{Shelly Vishwakarma\IEEEauthorrefmark{1}, Wenda Li\IEEEauthorrefmark{1}, Chong Tang\IEEEauthorrefmark{1}, Karl Woodbridge\IEEEauthorrefmark{2}, Raviraj Adve\IEEEauthorrefmark{3},  Kevin Chetty\IEEEauthorrefmark{1}}\\
    \IEEEauthorblockA{\IEEEauthorrefmark{1}Department of Security and Crime Science, University College London, UK}
    \IEEEauthorblockA{\IEEEauthorrefmark{2}Department of Electronic and Electrical Engineering, University College London, UK} 
    \IEEEauthorblockA{\IEEEauthorrefmark{3} Department of  Electrical and Computer Engineering, University of Toronto, Canada} \\
    \{s.vishwakarma, wenda.li,chong.tang.18, k.woodbridge,  k.chetty\}@ucl.ac.uk, 	
rsadve@comm.utoronto.ca
}
\maketitle
\begin{abstract}
This work presents an application of Integrated sensing and communication (ISAC) system for monitoring human activities directly related to healthcare. Real-time monitoring of humans can assist professionals in providing healthy living enabling technologies to ensure the health, safety, and well-being of people of all age groups. To enhance the human activity recognition performance of the ISAC system, we propose to use synthetic data generated through our human micro-Doppler simulator, SimHumalator to augment our limited measurement data. We generate a more realistic micro-Doppler signature dataset using a style-transfer neural network. The proposed network extracts environmental effects such as noise, multipath, and occlusions effects directly from the measurement data and transfers these features to our clean simulated signatures. This results in more realistic-looking signatures qualitatively and quantitatively. We use these enhanced signatures to augment our measurement data and observe an improvement in the classification performance by 5\% compared to no augmentation case. Further, we benchmark the data augmentation performance of the style transferred signatures with three other synthetic datasets- clean simulated spectrograms (no environmental effects), simulated data with added AWGN noise, and simulated data with GAN generated noise. The results indicate that style transferred simulated signatures well captures environmental factors more than any other synthetic dataset. 
  \end{abstract}
\providecommand{\keywords}[1]{\textbf{\emph{Keywords--}}#1}
\begin{IEEEkeywords}
WiFi based Sensing, ISAC, micro-Dopplers, human activity recognition, style transfer neural network, simulator, data augmentation
\end{IEEEkeywords}

\IEEEpeerreviewmaketitle

\section{Introduction}
\label{Sec:Intro}
Radio-frequency (RF) sensing is revolutionising commercial and consumer applications due to its vast usability, reliability, and affordability prospects. The technological advances in RF sensing have increased by leaps and bounds as it is minimally invasive, low cost, and privacy-preserving. It has been used for a wide range of applications, of specific interest here, human activity recognition has been the subject of intensive research in recent years \cite{erol2019gan,shah2019human,singh2019radhar,li2021semisupervised}. Humans are non-rigid bodies whose motion, when illuminated by RF signals, gives rise to frequency modulations. In addition, the relative movement of hands and legs give rise to additional Doppler returns popularly known as micro-Dopplers, which exhibit unique and discriminative features for different activities when observed in joint time-frequency space \cite{chen2003analysis,chen2006micro}. Over the last decade, RF sensors have used these micro-Doppler signatures to classify human activities for numerous applications ranging from law enforcement, security, and surveillance to ubiquitous sensing applications such as ambient assisted living and bio-medical applications \cite{shah2019rf,li2020passive}.

The emergence of these new applications in the consumer market has forced sensing systems to share the frequency bands with communication systems that were previously designed to operate separately and independently. As a result, the performance of both communication and sensing systems are susceptible to degradation due to the interference from neighboring wireless/sensing sources operating in the same band (treating one another as interferers) \cite{singh2017co}. Additionally, the unprecedented growth in the throughput demand for next-generation multimedia wireless services such as video streaming, data transfer amongst smart terminals (Internet of Things), high-dense deployment areas such as airports, and office spaces, has further congested the spectrum \cite{afaqui2016ieee,index2017global}. Therefore, there is an ongoing need to combine sensing and communication functionalities on the same wireless framework, popularly known as integrated sensing and communication (ISAC) \cite{restuccia2021ieee,tozlu2012wi,sharma2021passive,dang2020sensor}.  

The most crucial aspect of ISAC is designing a waveform capable of handling both sensing and communication functionalities simultaneously. The sensing waveform is mainly characterized by an ambiguity function, which provides insights into the proposed waveform's resolution capabilities (making it a popular tool for designing and analyzing sensing waveforms) \cite{richards2014fundamentals}. On the other hand, a communication waveform is designed by embedding and transmitting information, including components for synchronization, frequency offset estimation, channel estimation, and the data rate. In this regard, the conventional sensing waveforms have been reformed to embed communication information in the radar pulses to enable joint sensing, and communications \cite{dokhanchi2019mmwave}. However, when the sensing waveforms are optimized for ISAC applications, it only improves the sensing-centric performance, degrading the communication performance.
On the other hand, besides good communication performance, the IEEE 802.11 standard-compliant wireless waveforms have a favorable ambiguity function to perform additional functions such as detection and estimation of target parameters \cite{li2018wifi,duggal2020doppler}. Furthermore, these sub-6GHz WiFi signals have good wall penetration capabilities and contain richer multipath information. Therefore, they are being used to track and monitor indoor occupants for surveillance or healthcare purposes even through obstacles \cite{chetty2011through,tan2018exploiting,adib2015capturing}. However, due to limited bandwidth, these waveforms can not be used for fine localization; instead, the application is restricted to monitoring based on Doppler information.

In this work, we use IEEE 802.11g wireless transmissions at 2.462GHz from commercially available WiFi devices to recognize human activities directly related to healthcare applications ranging from everyday behaviors such as sitting and standing to falling over. We report on a passive ISAC system that does not transmit any signal on its own but instead relies on WiFi transmissions to detect and classify targets. The main idea is to use the existing standardized communication platforms to implement an augmented sensing system (without alterations to the existing system). Since this approach neither requires excessive costs nor incurs privacy concerns, it is considered a promising alternative to conventional wearable sensor methods. 

We gather returns from five different human subjects undergoing ten different activities using our passive ISAC system comprising of a WiFi AP, two antennas, and two national instruments (NI) USRP-2921 \cite{USRP2921}. These measurements act as our baseline to train a neural network. Specifically, we process the returns to generate the corresponding micro-Doppler signatures to be used to train a neural network. A deep learning framework jointly learns informative features and classification boundaries without using an additional feature extraction algorithm. Unfortunately, training any deep learning model requires a vast amount of good quality labeled data, which is hard to gather in practice, especially the falling over motions. In this case, the deep learning network can easily over-fit and lose the generalization capability to recognize unseen samples. Moreover, the data is affected by various environmental conditions, sensor parameters, and target characteristics, affecting the performance of deep learning algorithms. 
\begin{figure*}[htbp]
\centering
\includegraphics[scale=0.6]{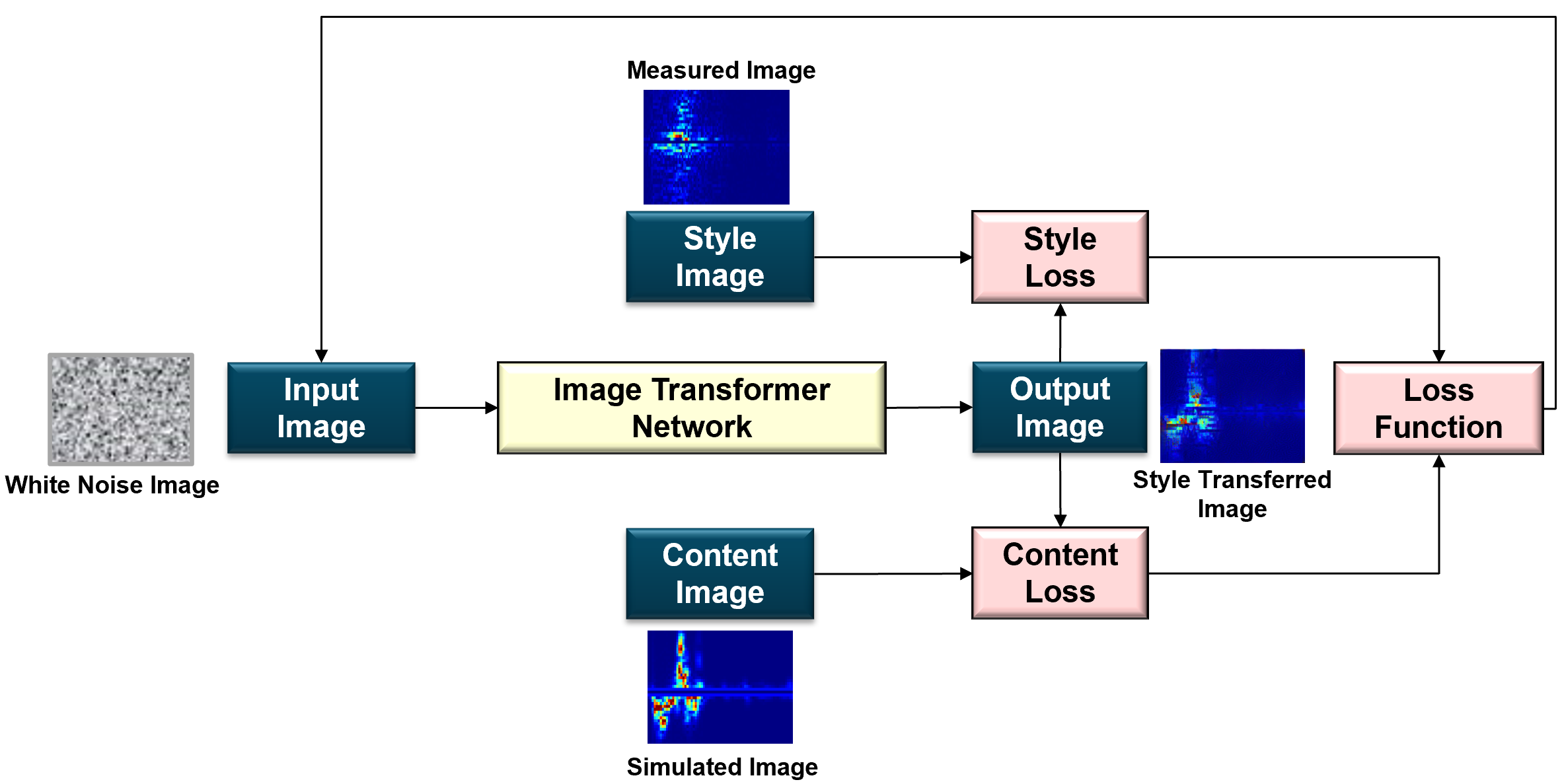}
\caption{Neural style transfer: an image transformer framework}
\label{fig:styleframeworkoverview}
\end{figure*}
Unlike the fields of vision and image processing, we cannot
simply augment data by performing cropping, rotation, and flipping operations as it might distort the kinematic fidelity of the signatures. Moreover, the sensing community has limited access to open databases that contain large volumes of experimental data. In response to open science practices towards accelerating, improving, and accumulating scientific knowledge for the research community to re-use and build upon, we have publicly released a motion capture driven radar simulator, \textbf{SimHumalator}, \url{https://uwsl.co.uk/}. In doing so, we built a user-friendly graphic user interface (GUI) to wrap around the simulator \cite{vishwakarma2021simhumalator}. In addition, we wrote a user guide detailing how to operate the software while clearly explaining the complexities of the signal processing techniques employed. The simulator can alleviate the well-known cold-start problem in sensing where there is a lack of usable data to train deep learning networks. By imitating the operation of a real-world system under different operating conditions and simulating the corner cases that are hard to reproduce in practice, it serves to generate large volumes of training data and reduce the labor and expense involved in data collection. 

In our previous work \cite{tang2021augmenting}, we leveraged human micro-Doppler data generated using SimHumalator to augment our measurement data. The results highlight that the classification performance can be improved for cases in which only limited experimental data is available for training. However, the spectrograms considered in this study did not account for environmental factors such as noise, multipath, and propagation loss, resulting in very clean simulated spectrograms. 

In \cite{vishwakarma2021gan}, we further investigated the gains from data augmentation using a more realistic training dataset, with two types of noise added to the SimHumalator data.  The first was additive white Gaussian noise (AWGN). However, AWGN is pixel-independent, spatially uncorrelated, and cannot incorporate environmental effects such as clutter, multipath, or occlusions. We, therefore, proposed a second noise modeling framework based on generative adversarial networks (GANs) to mimic more complex real-world scenarios. The adversarial training learned the noise distribution model directly from the measured spectrograms. However, the limitation of this approach was that it used only the non-activity zones in the measured spectrograms to extract the noise parameters. Therefore, it does not account for multipath and occlusion effects arising from target interaction with the environment. 

This work investigates the possibility of extracting noise, multipath, and occlusions effects directly from real spectrograms using an image transformer network called neural style transfer \cite{kotovenko2019content}. The style transfer framework has been extensively applied in image stylization \cite{sanakoyeu2018style} and texture synthesis \cite{risser2017stable,elad2017style}. A neural network-based style transfer framework requires two images: a content image and a style image. It separates the content of an image from its style and tries to recast that content in the style of another image (as shown in Fig. \ref{fig:styleframeworkoverview}). In other words, the network combines the style (or texture information) of one image and content of the other image to generate a new image by extracting appropriate feature responses from image transformer network layers. This objective is achieved by minimizing a loss function that is the sum of two separate loss terms- content loss and style loss. The content loss represents the dissimilarity between the content image and the output image, whereas the style loss represents the dissimilarity between the style image and the output image. 

In order to generate realistic synthetic micro-Doppler signatures, we consider our clean simulated spectrogram generated from SimHumalator to be the content image and our measured spectrogram to be the style image. The resultant output image from the style transfer framework captures the micro-Doppler content and the envelope from the simulated content image. In contrast, the target-dependent multipath, clutter, noise, and occlusion effects are transferred as style features onto the output image thereby capturing the characteristics of the measured signatures. 

We qualitatively and quantitatively compare the measured signatures with synthesized signatures generated under different scenarios such as no noise case (clean simulated signatures), signatures with added AWGN noise, signatures with added GAN generated noise, and style transferred signatures. To extract the local features across both the measured and synthesized signatures, we use the speeded-up robust features (SURF) algorithm \cite{ping2013review}. Importantly, as compared to the other approaches to generate synthetic signatures, we observe that the features extracted using the style-transferred signatures better overlap with features extracted from the measured signatures. 

We report on a detailed augmentation study using the four simulation datasets- one with no noise, second with added AWGN noise, third with GAN-based noise, and finally with data generated through style transferred framework. 

To sum up, our contributions in this paper are the following:
\begin{enumerate}
    \item We develop a style transfer neural network to generate more realistic synthetic signatures that overlap in feature space, with actual measurement data. The synthesized signatures now include propagation effects such as multipath, clutter, and occlusion without running complex electromagnetic simulations. 
    \item We qualitatively and quantitatively characterize the synthesized signatures relative to measured signatures and compare the performance with respect to the-no noise, signatures with AWGN noise, and signatures with GAN generated noise cases. 
    \item We use the synthetic signatures generated for data augmentation to solve the practical problem associated with limited or unbalanced micro-Doppler training datasets.
    \item Finally, we study two data augmentation methods: Replacement and Augmentation and validate their performance in different scenarios using a simple deep convolutional neural network. 
\end{enumerate}

Our paper is organized as follows. Section \ref{sec:exp_setup} describes the experimental setup and data collection using two synchronized systems- measurement ISAC system and a Kinect-v2 motion capture system. Section \ref{sec:style_transfer} describes the proposed style transfer framework for more realistic synthetic data generation. This section also demonstrates the qualitative and quantitative comparison of synthetic signatures relative to measured signatures. Section \ref{sec:meas_simulations_aug} shows interesting classification scenarios under the four mentioned cases. We finally conclude our paper in Section \ref{sec:conclusion}.
\begin{figure}[htbp]
\centering
\includegraphics[scale=0.40]{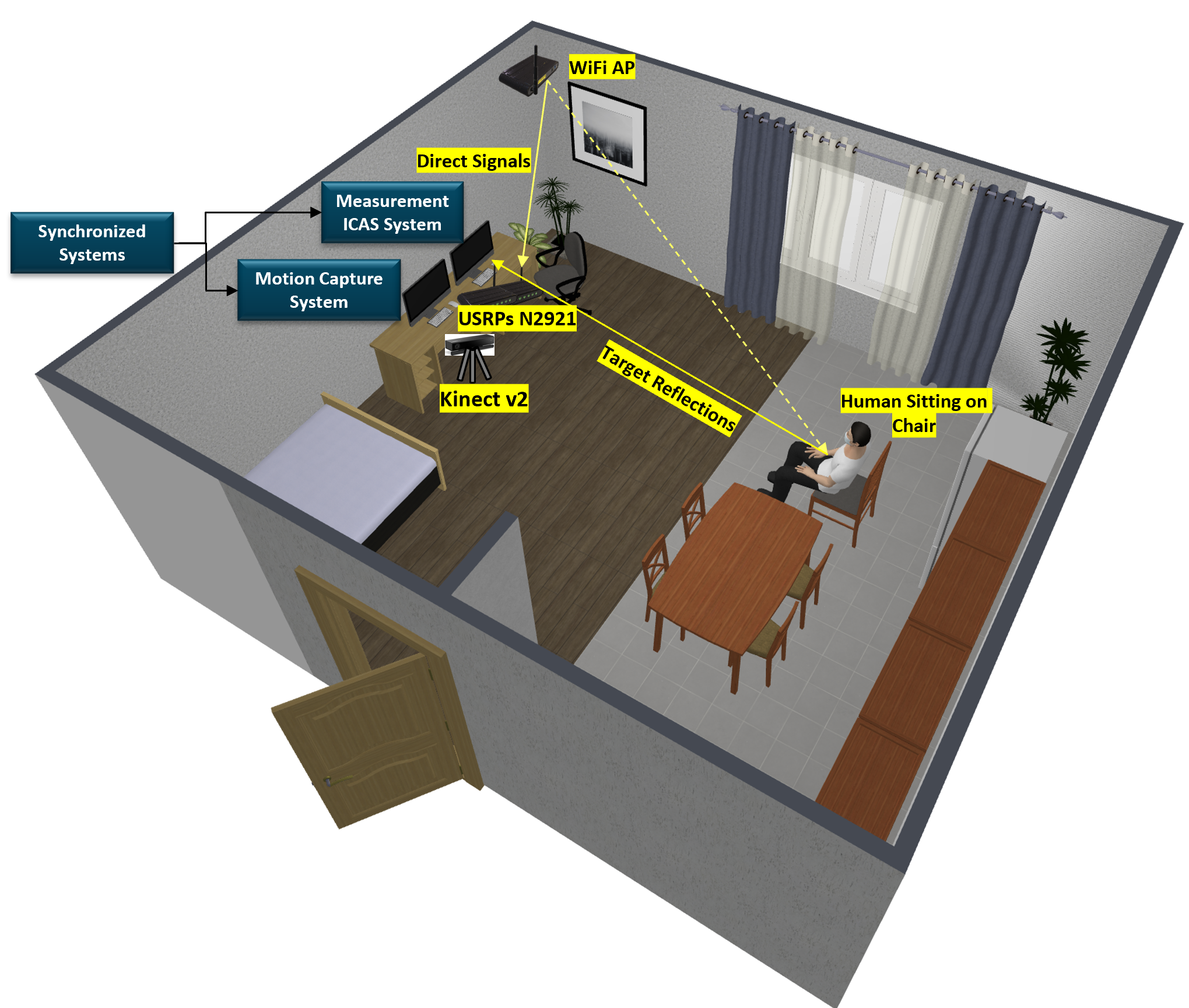}
\caption{Experimental Setup comprising of two synchronised systems- a passive ISAC measurement system and a Kinect-v2 motion capture system to gather the human animation data to be subsequently used for generating simulated signatures.}
\label{fig:experimentalenvironment}
\end{figure}
\section{Experimental Setup And Data Collection}
\label{sec:exp_setup}
Fig. \ref{fig:experimentalenvironment} presents our experimental setup comprising two synchronized systems: an infrared motion capture Kinect v2 sensor and a non-contact human activity monitoring passive ISAC system. This section briefly describes the passive ISAC system used for measurement micro-Doppler data capture, our simulation software, SimHumalator, used for generating the corresponding synthetic signatures. 
\subsection{Passive ISAC System}
The passive ISAC system shown in Fig. \ref{fig:experimentalenvironment} is set up using two National Instruments (NI) USRP-2921 \cite{USRP2921}, two Yagi antennas, each with a gain of 14dBm, and a Netgear R6300 transmitter acting as the WiFi AP. The system uses one antenna as a reference antenna to capture direct WiFi transmissions at a center frequency of $2.472$GHz from the AP. Simultaneously, it uses a second antenna as a surveillance antenna to capture signals reflected off-targets in the propagation environment. The signals received at both reference and surveillance antennas are cross-correlated to generate the target signatures in real-time. Since the signal bandwidth for IEEE 802.11g WiFi transmissions is limited to 20MHz (insufficient to locate targets in most indoor scenarios), we extract only the time-varying micro-Doppler information in joint time-frequency space, also known as spectrograms. 
\subsection{SimHumalator}
In the experiments, we synchronize and co-locate both the ISAC system and the Kinect v2 sensor to ensure that the animation and measured data express identical
motion information. Kinect captures the three-dimensional time-varying skeleton information of the dynamic human subject. We use this time-varying skeleton information as an input to our open-source simulation tool SimHumalator to generate human micro-Doppler signatures \cite{vishwakarma2021simhumalator}. To mimic realistic wireless transmissions, SimHumalator generates a IEEE 802.11g standard-compliant WiFi signal using MATLAB's WLAN toolbox and combines it with animation data to simulate the reflected signals. Interested readers can download the simulator and read the detailed working methodology from \url{https://uwsl.co.uk/}.

\subsection{Data Collection}
We monitor five participants- (two males and one female) performing the first six activities and (two other males and same female) performing the remaining four activities described in Table \ref{table:Activity_Description}.  We restrict our measurements to direct line-of-sight conditions with human subjects moving between 0.8m to 3.8m in front of the system to mimic distances in a typical indoor environment. We record each activity for a duration of 5-10sec depending upon the nature of the activity. We repeat these measurements 20 times for each participant, resulting in 60 measurements for each activity. Overall, we gather 600 measurement data and data from 600 simulation over ten activities. We use two synchronized systems to demonstrate the same motion characteristics shared by the two systems. However, in practice, we do not use synchronized data for our further studies. 
\begin{table}[t]
\centering
\caption{Human Activity Description}
\label{table:Activity_Description}
\begin{tabular}{|
>{\columncolor[HTML]{EAF8E0}}c |c|}
\hline
Name        & \cellcolor[HTML]{EAF8E0}Activity               \\ \hline
Activity 1  & \cellcolor[HTML]{FFFFFF}Sit down on chair      \\ \hline
Activity 2  & Stand up from chair                            \\ \hline
Activity 3  & Stand up from chair to walk                    \\ \hline
Activity 4  & Walk to sit down on chair                      \\ \hline
Activity 5  & Walk to fall                                   \\ \hline
Activity 6  & Stand up from ground to walk                   \\ \hline
Activity 7  & Bodyrotating                                   \\ \hline
Activity 8  & Walking back and forth                         \\ \hline
Activity 9  & Punching                                       \\ \hline
Activity 10 & Pickup object from ground and dropping it back \\ \hline
\end{tabular}
\end{table}
Note that the spectrograms generated through SimHumalator only contain the target's motion information; specifically these signatures do not capture environmental factors such as noise, propagation loss, occlusions, and multipath. Therefore, in the subsequent section, we present our proposed style transfer framework to generate micro-Doppler signatures extracting all the environmental effects straight from the measurement data. 

\section{Style transfer: Realistic Synthetic Database Generation and Evaluation}
\label{sec:style_transfer}
Neural networks are effective feature extractors. They generally comprise of multiple layers of
simple computational units that process information in a feed-forward manner. Each layer consists of a collection of filters that extracts a unique feature from the input image. Thus, the output of a given layer consists of feature maps that are differently filtered versions of the input image.

In this framework, we use one such deep neural network, VGG-19 \cite{canziani2016analysis}, to extract the content or motion characteristics of the simulated micro-Doppler signature and style or environmental effects in the measured micro-Doppler signature to generate the third signature, the combination of two.  Since this framework is only used for feature extraction at different layers, we remove all the fully connected layers otherwise used for classification purposes. The resulting network uses 38-layered architecture. 
\subsection{Micro-Doppler Style Transfer Framework}
We consider our simulated micro-Doppler signature ($c$) as the content image and measured micro-Doppler signature ($s$) as the style image. We aim to generate a third image that retains the motion information of the simulated signature and acquires environmental effects such as clutter, multipath, and noise as a background from the measured signature. 
\begin{figure*}[htbp]
\centering
\includegraphics[scale=0.7]{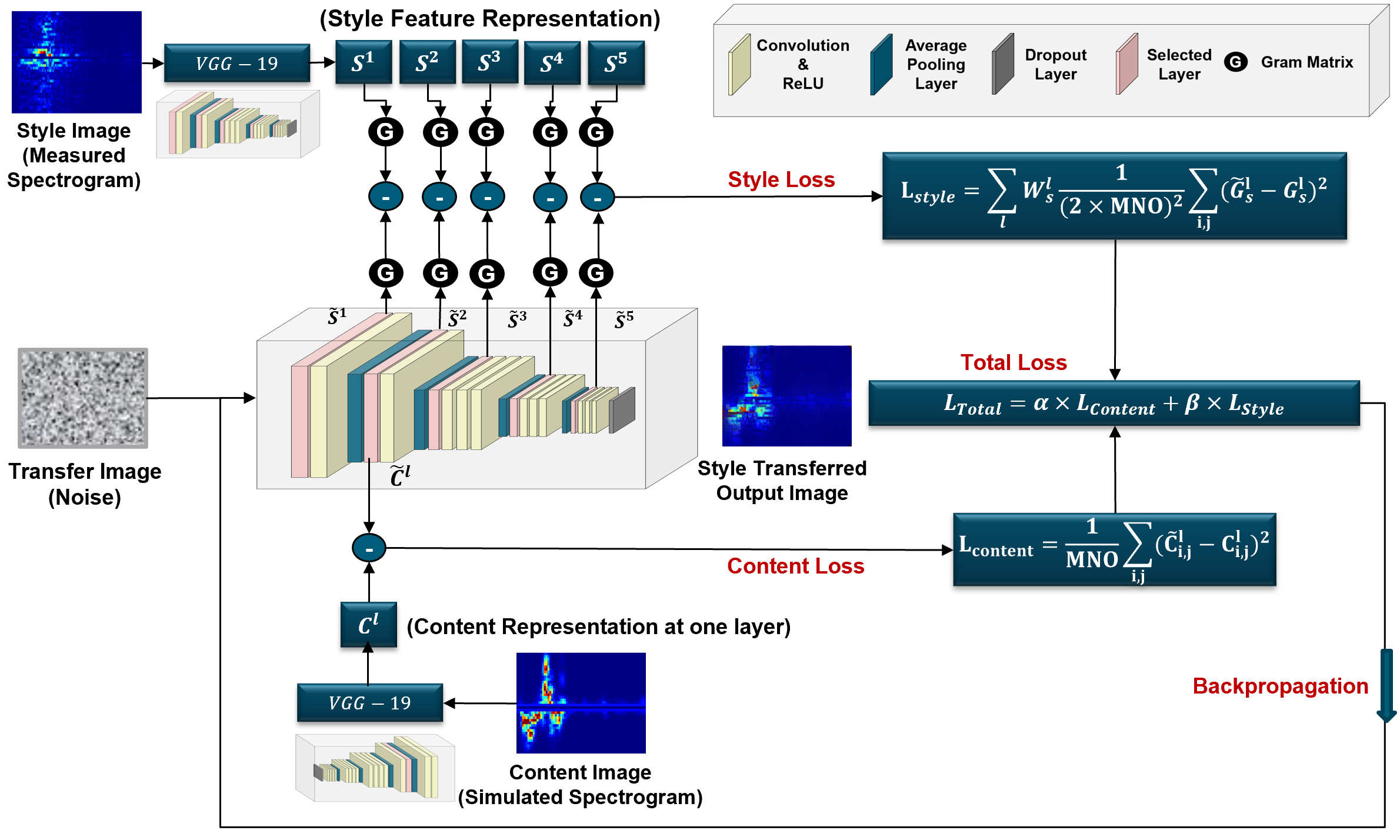}
\caption{The network overview of style transfer framework}
\label{fig:styleframework}
\end{figure*}
\subsubsection{Content Features Extraction}
In order to capture the simulated signature's content information encoded at a different convolutional layer of the VGG-19 network, we run the simulated micro-Doppler image $c$ through the network and gather feature maps generated at each layer in feature matrix $C_{feat}^{l} \in [N^l\times D^l]$. Note that each layer $l$ in the neural network has $N^l$ different filters to generate the $N^l$ individual filter responses each of size $D^l$. Here, $D^l$ is the product of height $M$, width $N$, and channel $O$ of each feature map. Concurrently, we pass a white noise image $\tilde{c}$ through the network and compute its feature maps at every layer and store them in a matrix $\tilde{C}_{feat}^{l} \in [N^l\times D^l]$. 

To update the initial white noise image $\tilde{c}$ with the content information of the content image $c$, we minimize the following loss function and compute its gradient using standard error back-propagation iteratively. 
\begin{equation}
\label{eq:contentloss}
    L_{content}=\sum_{l}\frac{1}{MNO}\sum _{i,j}(\tilde{C}^{l}_{feat}-C^{l}_{feat})^2
\end{equation}
The loss is computed as sum of the Euclidean distance between the activation of the content image $C_{feat}^{l}$ and the activation of the output image $\tilde{C}_{feat}^{l}$ at each layer. Note that $i$ denotes the filter number, $j$ denotes the position in the $i^{th}$ filter. The process is repeated until it generates the same response in a certain layer of the CNN as the original content image. The objective of content loss is to make the features of the output image match the features of the content image. 

\subsubsection{Style Features Extraction}
Unlike the content features, the style features are captured by computing correlations between the different filter responses at each layer. These feature correlations are represented by the Gram matrix 
\begin{equation}
\label{eq:grammatrix}
   G^l_s=\sum _{k}S_{i,k}^{l}S_{j,k}^{l}
\end{equation}
Effectively, $G^l_s$ is the inner product between the vectorized feature maps $i$ and
$j$ at layer $l$. 

We consider our measured signatures as the style image and a white noise image as the initial input image updated continuously to extract the style features. Therefore, the objective of style loss is to make the texture of the output image match the texture of the measured signatures by minimizing the objective function 
\begin{equation}
\label{eq:styleloss}
   L_{style}=\sum_{l}W^l_s\frac{1}{MNO}\sum _{i,j}(G^l_s-\tilde{G}^l_s)^2
\end{equation}
$L_{style}$ is computed as the weighted ($W^l_s$) sum of the mean squared difference between the Gram matrix of the style image and the Gram matrix of the output image at different layers. We believe that by correlating features from multiple layers, we can capture the environmental factors directly from the measured signatures while ignoring the global arrangements of motion characteristics.
\begin{figure}[htbp]
\centering
\includegraphics[scale=0.55]{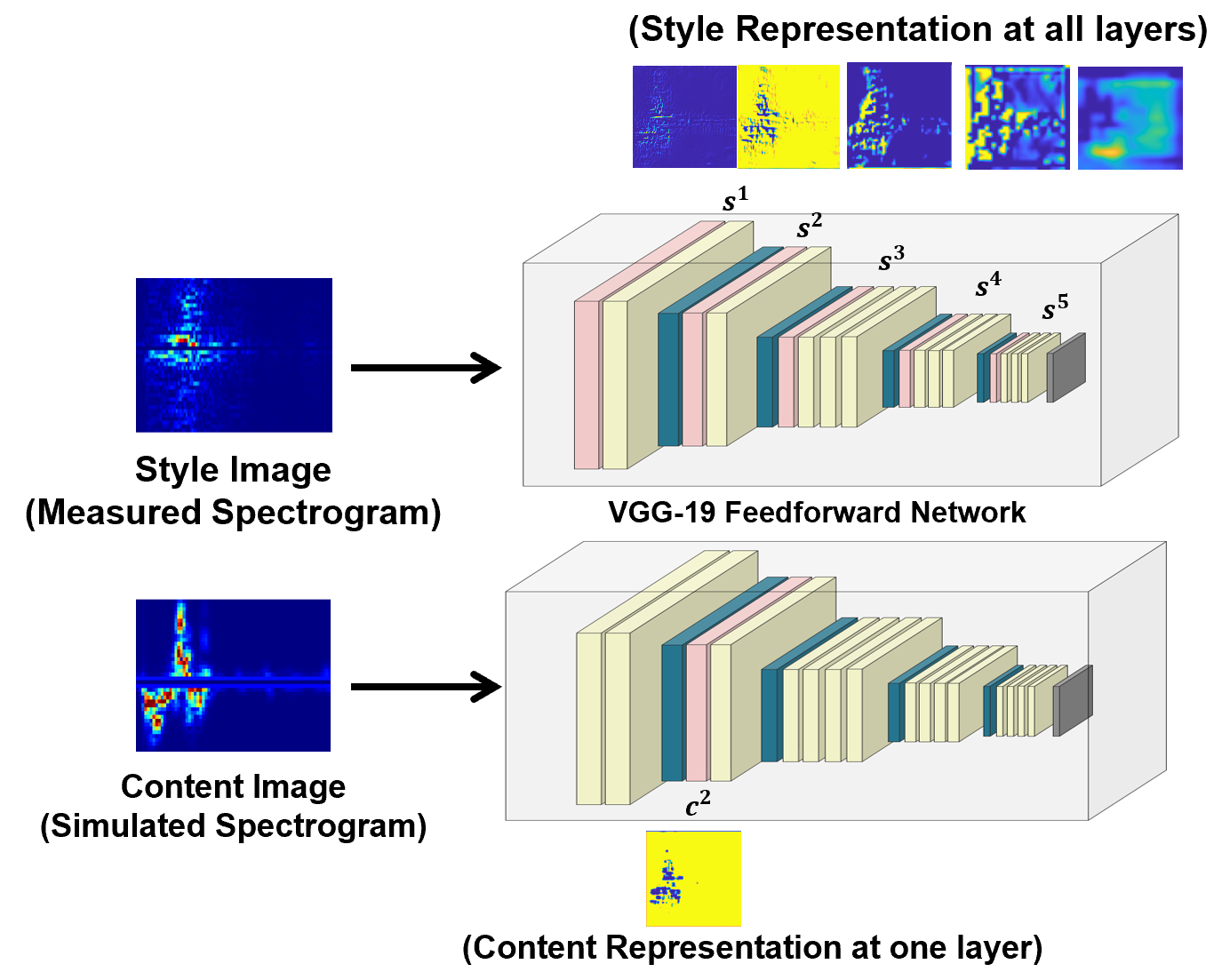}
\caption{Style feature visualisations on layer $conv1_1$, $conv2_1$, $conv3_1$, $conv4_1$ , and $conv5_1$ and content feature map visulation at $conv2_1$.}
\label{fig:style_network}
\end{figure}
\subsubsection{Style Transfer Algorithm}
Fig. \ref{fig:styleframework} presents our proposed micro-Doppler style transfer framework. We initially begin with a white noise image as our transfer image. We pass this image through the neural network to extract its style features and the content features over different layers and compute the style loss and the content loss between this image and the style image (measured signatures) and content image (simulated signatures), respectively. We repeatedly optimize the transfer image to combine the style of the measured signature and the content of the simulated signatures by back-propagating the gradient. In order to obtain a good transfer image, we jointly minimize the loss function $L_{total}$, a weighted combination of content loss $L_{content}$ and style loss $L_{style}$; $\alpha$ and $\beta$ denote weighting factors for content loss and style loss, respectively.
\begin{equation}
\label{eq:totalloss}
   L_{total}=\alpha L_{content}+ \beta L_{style}
\end{equation}

We note that combining the content of one image with the style of another does not usually guarantee that the output image will match both constraints simultaneously.  A strong emphasis on style will result in images that match the appearance of the measured signatures but hardly show any of the motion content. When placing a strong emphasis on content, one can identify the motion characteristics, but the background of the measured signature is not well-captured. Therefore, we minimize the combined loss function for both content and style to emphasize reconstructing a combination of two.
\begin{figure}[htbp]
\centering
\includegraphics[scale=0.5]{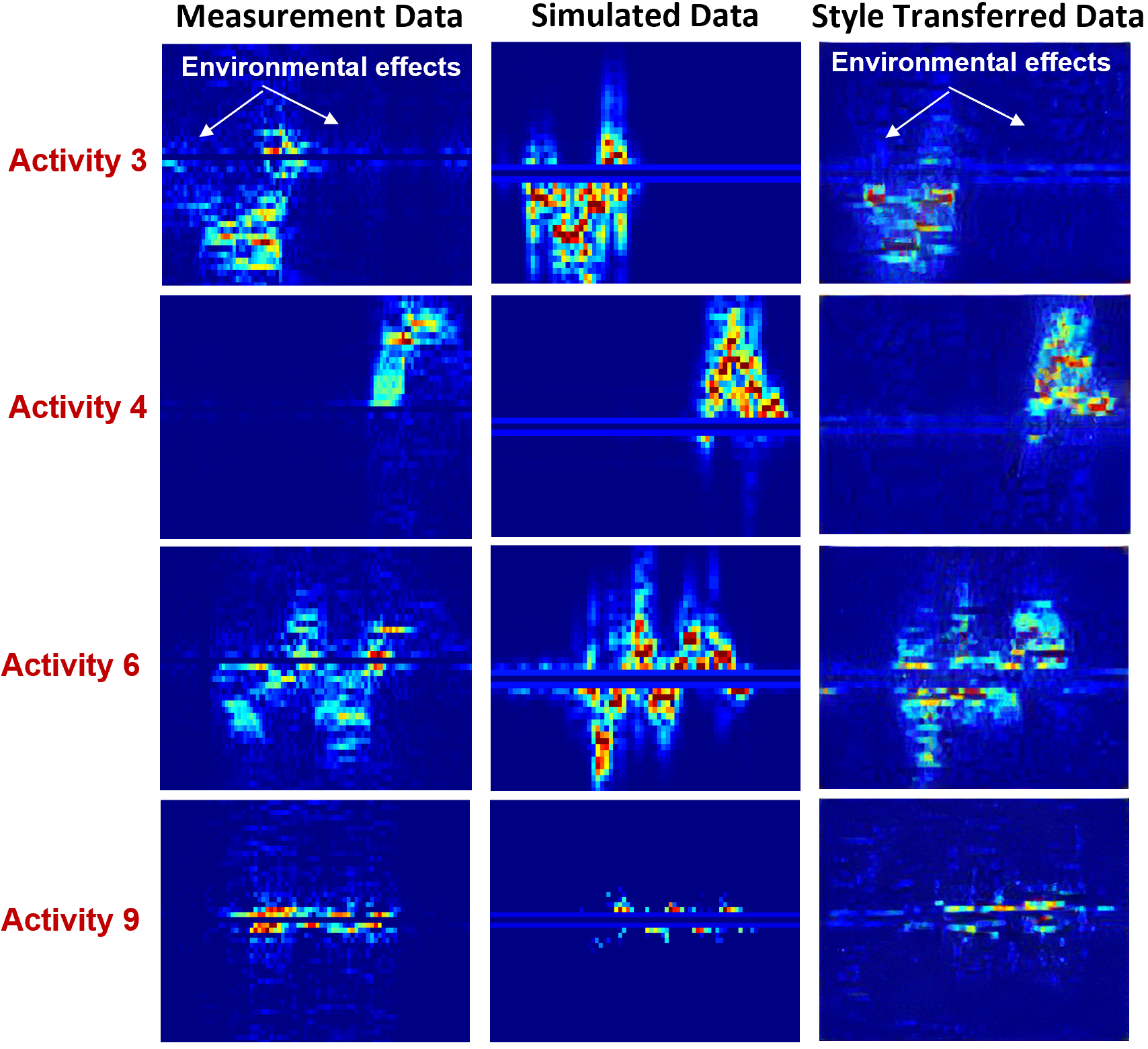}
\caption{Examples of the measured signatures, corresponding clean signatures and style transferred signatures.}
\label{fig:spectrogram_comparison_orig}
\end{figure}
Fig. \ref{fig:style_network} presents the five style visualisations on layer $conv11$, $conv21$, $conv31$, $conv41$ , and $conv51$. We can observe that the reconstructions from the style features at different layers produce textured versions of the content image. We use the second convolution layer  $conv21$ as the content feature extraction layer as the VGG-19 network can extract the motion characteristics effectively from lower layers. 

We pass each simulated spectrogram through the style transfer framework, optimize the total loss for $2500$ iterations, and keep the ratio of $\alpha$ to $\beta$ to be 1e-3. We ran our algorithm using Matlab 2020b, where all the variables are stored as 64-bit floats, with system configuration specified as- Intel(R) Core(TM) i7-10750H CPU running at 2.60GHz, 2592 Mhz, 6 Core(s), 12 Logical Processor(s). 

Fig.\ref{fig:spectrogram_comparison_orig} presents some spectrogram examples generated through the proposed framework and its corresponding measured and clean simulated spectrogram pairs. Effectively, the synthesized transfer images are rendered in the style of the measured signatures capturing all the environmental factors and keeping the motion content the same as that of the simulated signatures.    

\begin{figure*}[htbp]
\centering
\includegraphics[scale=0.6]{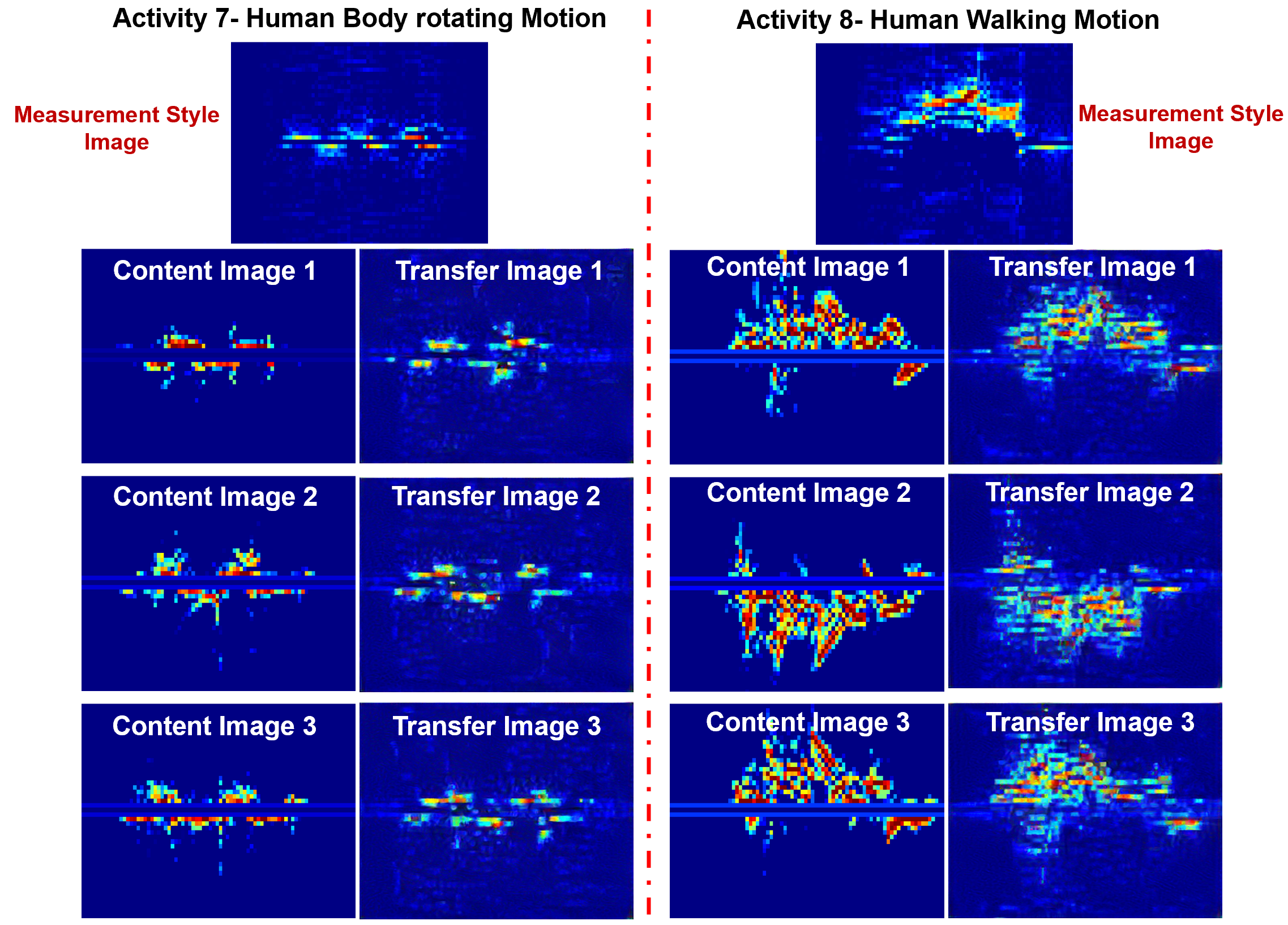}
\caption{Examples of transfer of style using a single measurement image. Single measurement image, single clean simulated content image is passed through the network. For other simulated signatures, the same measurement image is used as the style image. Generating the whole database requires multiple passes of multiple images through the network.}
\label{fig:spectrogram_comparison_orig1}
\end{figure*}

Note that, for a particular activity, we pick only one measured signature as our style image and transfer its texture to all the simulated signatures for that activity one by one. Fig. \ref{fig:spectrogram_comparison_orig1} presents two such examples where we use one measured image for activity 7 and one from activity 8.  We transfer their background to three synthesized signatures belonging to the corresponding activity. We repeat this exercise for all the spectrograms in each of the ten activities. 

\begin{figure*}[htbp]
\centering
\includegraphics[scale=0.72]{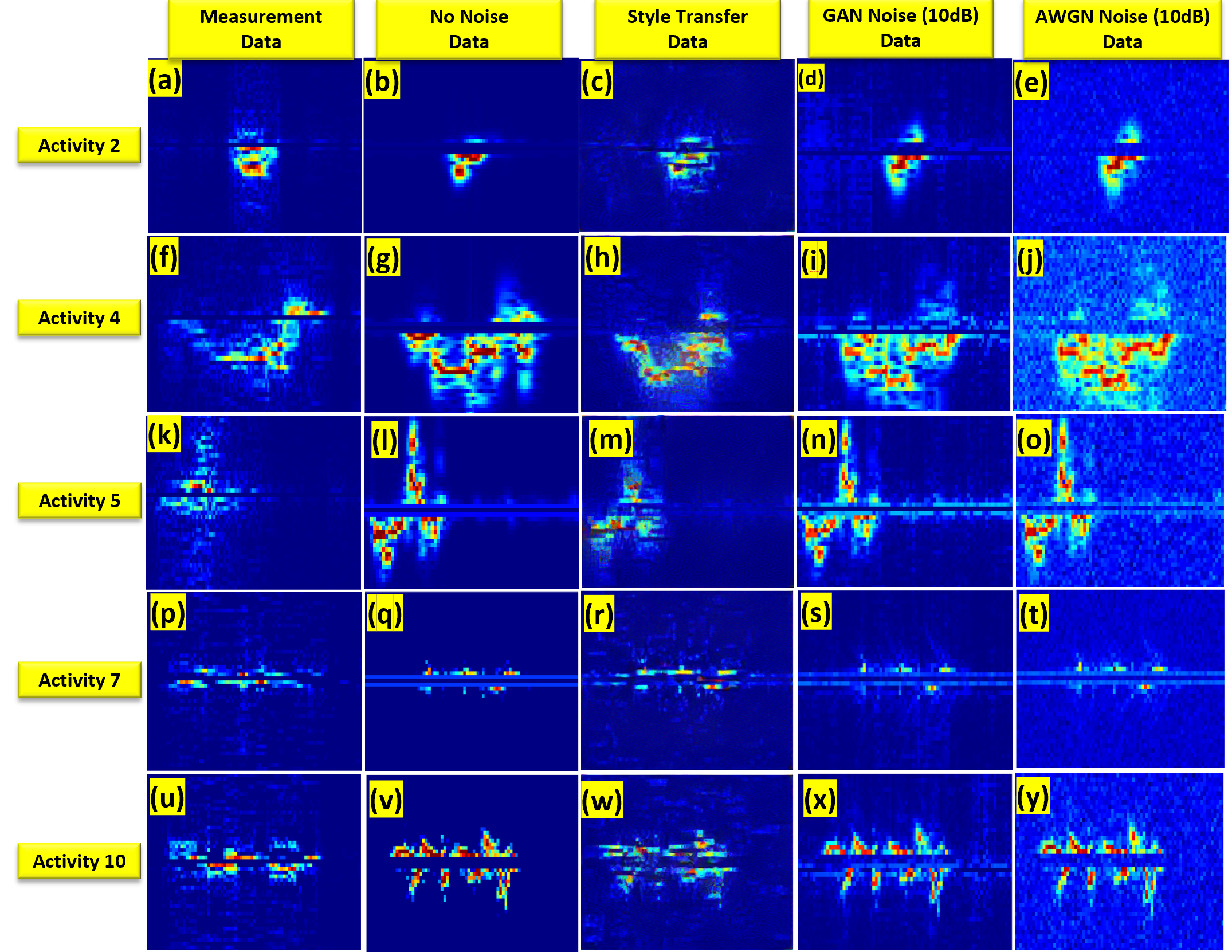}
\caption{Qualitative Analysis: (a)-(e) Measured micro-Doppler spectrograms, Clean simulated spectrograms, Style transferred spectrograms, GAN noise added spectrograms, and AWGN noise added spectrograms of a human undergoing Activity 2. (f)-(j) spectrograms of a human undergoing Activity 4, (k)-(o) spectrograms of a human undergoing Activity 5, (p)-(t) spectrograms of a human undergoing Activity 7 and (u)-(y) spectrograms of a human undergoing Activity 10 respectively.}
\label{fig:spectrogram_comparison}
\end{figure*}
\subsection{Qualitative Analysis: Spectrograms Visual Evaluation}
Fig. \ref{fig:spectrogram_comparison} presents a qualitative comparison of synthesized and the measured micro-Doppler spectrograms of a human undergoing five activities, including- Activity 2 (standing up motion), Activity 4 (walk to sit down on a chair), Activity 5 (walk to fall), Activity 7 (body rotation), and Activity 10 (picking up an object from the ground and dropping it back).

The qualitative similarity can be observed between measured signatures in the first column and the clean simulated signatures in the second column. However, as our simulations do not account for various environmental factors, the spectrograms are very clean relative to the measured spectrograms. Column 3, Fig. \ref{fig:spectrogram_comparison}(c),(h), (m), (r) and (w) presents the style transferred spectrograms. As can be seen from the figures, the target-dependent multipath, clutter, and occlusion effects are captured quite well from the measured spectrograms. The effects are more prominent around the signatures just like the measured spectrograms and can also capture shadowing to a greater extent, with some parts of the signatures now has less intensity, unlike the clean spectrograms. 

We synthesize two other signatures, one with GAN generated noise in Fig. \ref{fig:spectrogram_comparison} column 4 and other with AWGN noise at 10dB signal to noise ratio (SNR) in Fig. \ref{fig:spectrogram_comparison} column 5. As GAN-generated signatures capture the noise from non-activity zones of the measurement spectrogram, they cannot capture target-dependent environment effects well. In summary, from all the synthesized signatures presented in Fig. \ref{fig:spectrogram_comparison}, style transferred signatures best capture the noise distribution of the measured spectrograms and hence are more realistic than any other case. 

\subsection{Quantitative Analysis: Feature Space Visualisation}
\begin{figure*}[htbp]
\centering
\includegraphics[scale=0.6]{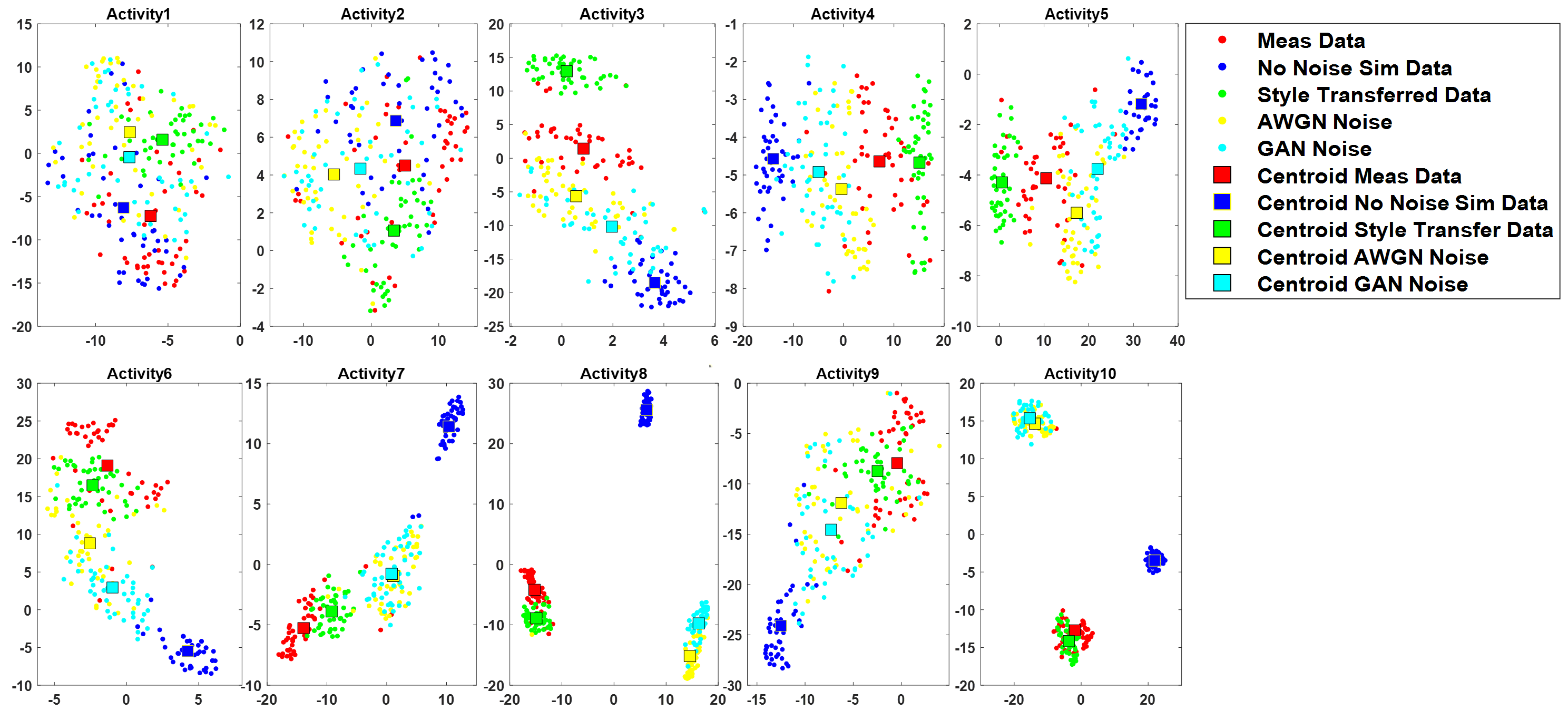}
\caption{Quantitative Analysis: Scatter plots for measured and synthesized signatures using SURF features.}
\label{fig:tsne}
\end{figure*}

To understand this further, we made a quantitative comparison of both the measured and the synthesized spectrogram sets based on the SURF features for all the ten activities \cite{murillo2007surf}. We visualize these features in a two-dimensional space using the t-distributed Stochastic Neighbor Embedding (t-SNE) technique \cite{van2008visualizing}, and show the corresponding scatter plots in Fig.\ref{fig:tsne}.  The extracted features represent essential and unique attributes of the signatures.
Ideally, if the signatures do not share spatial and temporal similarities, their features should be clustered sparsely/randomly (like Activity 1). On the other hand, if the signatures share the same features space, they should be clustered together in the scatter plots like Activity 2 to Activity 10. 
 
As can be seen from Fig. \ref{fig:tsne}, for Activity 1 and Activity 2, the feature points for almost all the datasets, including- measured and synthesized signatures share the same latent space. The remaining eight activities from Activity 3 to Activity 10 have well-clustered data points. As expected from our qualitative evaluation of the signatures, the style transferred signature's feature points form the cluster closest to the measurement data cluster for most activities, especially Activity 6, 7, 8, and 10. To further support our analysis, we compute the centroids of each cluster using the k-means clustering algorithm and compute the Euclidean distance of the measurement data centroid relative to other synthesized dataset centroids for all the activities \cite{likas2003global}. 

We report the results in Table \ref{table:QuantitativeAnalysis}.
\begin{table*}[htbp]
\centering
\caption{Quantitative Analysis: Euclidean distance of synthesized signatures dataset cluster centroid to measurement data centroid.}
\label{table:QuantitativeAnalysis}
\begin{tabular}{|
>{\columncolor[HTML]{EAF8E0}}c |c|c|c|c|}
\hline
\begin{tabular}[c]{@{}c@{}}Activity\textbackslash{}Euclidean Distance \\ (Between Centroid of Meas Data Features and)\end{tabular} & \cellcolor[HTML]{EAF8E0}No Noise Sim Data & \cellcolor[HTML]{EAF8E0}Style Transferred Sim Data & \cellcolor[HTML]{EAF8E0}Added AWGN Noise & \cellcolor[HTML]{EAF8E0}Added GAN Noise \\ \hline
Activity 1                                                                                                                                & \cellcolor[HTML]{FFFFFF}\textbf{1.8}      & \cellcolor[HTML]{FFFFFF}8.7                        & 10.1                                     & 7                                       \\ \hline
Activity 2                                                                                                                                & \textbf{2.5}                              & 4.1                                                & 10.2                                     & 6.6                                     \\ \hline
Activity 3                                                                                                                                & 20.8                                      & 12.4                                               & \textbf{7.9}                             & 12.6                                    \\ \hline
Activity 4                                                                                                                                & 20.1                                      & \textbf{7.3}                                       & 7.3                                      & 11.7                                    \\ \hline
Activity 5                                                                                                                                & 20.3                                      & 9.7                                                & \textbf{6}                               & 10.6                                    \\ \hline
Activity 6                                                                                                                                & 26.2                                      & \textbf{2.6}                                       & 9.9                                      & 15.9                                    \\ \hline
Activity 7                                                                                                                                & 47.4                                      & \textbf{5.6}                                       & 19.8                                     & 19.4                                    \\ \hline
Activity 8                                                                                                                                & 56.3                                      & \textbf{4.8}                                       & 23.5                                     & 28.8                                    \\ \hline
Activity 9                                                                                                                                & 20.6                                      & \textbf{2.4}                                       & 7.6                                      & 9.9                                     \\ \hline
Activity 10                                                                                                                               & 40.6                                      & \textbf{2.1}                                       & 33.2                                     & 34.6                                    \\ \hline
\cellcolor[HTML]{EAF8E0}Mean                                                                                                              & \cellcolor[HTML]{FFFFC7}25.7              & \cellcolor[HTML]{FFFFC7}\textbf{5.9}               & \cellcolor[HTML]{FFFFC7}13.6             & \cellcolor[HTML]{FFFFC7}15.7            \\ \hline
\end{tabular}
\end{table*}
We observe that the clean data has the highest distance from the measurement data, mainly because it does not include crucial factors arising in any natural environment such as noise, clutter, multipath, and other target-dependent phenomena. To some extent, the synthesized data using AWGN and GAN-generated noise can bridge the gap with measurement data resulting in lower distances, clearly evident from the results. However, the environmental factors introduced are not adequate. Crucially, the style transfer data shows the lowest mean distance from the measurement data in the feature space, indicating that these signatures can capture the environmental factors into the signatures very well and should serve as good signatures to augment an otherwise limited measurement dataset. 

Overall, our quantitative results are consistent with our qualitative analysis indicating the feasibility and robustness of the micro-Doppler style transfer framework. In the next section, we use these signatures to augment our measurement signatures to train a neural network and investigate the resulting classification performance. The crucial benefit of using synthetic data is that we can generate a significant amount of training data quickly. 
 
\section{Experimental Classification Results and Analyses}
\label{sec:meas_simulations_aug}
To investigate the classification performance under various experimental scenarios, we designed an 8-layered convolutional neural network. We reshaped the size of our input spectrograms to be $100 \times 100$. The learning rate of the adaptive moment estimation optimizer is set to $0.001$; the batch size to $64$; the output shape is modified to $10$ for our multi-class classification task, the loss function is categorical cross-entropy; the epoch is set to $100$.

We keep identical training parameters to investigate the classification performance in the following five scenarios.
\begin{itemize}
    \item \textbf{Case 1:} Both training and test dataset comprise measurement data only (TMTM).
    \item \textbf{Data Augmentation Scenario}
    \begin{itemize}
    \item \textbf{Case 2:} Measurement training dataset augmented using the style transferred simulation data.
    \item \textbf{Performance Benchmarking Across Multiple Scenarios}
    \begin{itemize}
    \item \textbf{Case 3:} Measurement training dataset augmented using the no noise clean simulation data.
    \item \textbf{Case 4:} Measurement training dataset augmented using the AWGN noise added simulation data.
    \item \textbf{Case 5:} Measurement training dataset augmented using the GAN noise added simulation data.
    \end{itemize}
    \end{itemize}
\end{itemize}

\subsection{Case 1: Train and Test With Measurement Data Only (TMTM)}
To begin with, we split our total measurement dataset ($M=600$) into two 50\% for training and 50\% for the test. Note that only the unseen measurement data is used to test the network's classification performance. Table \ref{table:TMTM} reports the classification accuracy for case 1 in form of a confusion matrix. 
\begin{table*}[]
\centering
\caption{Confusion matrix for TMTM case: 50\% dataset used for training and remaining 50\% used for testing the performance of the network.}
\label{table:TMTM}
\begin{tabular}{|
>{\columncolor[HTML]{EAF8E0}}c |c|c|c|c|c|c|c|c|c|c|}
\hline
True Class/Predicted Class & \cellcolor[HTML]{EAF8E0}Activity1 & \cellcolor[HTML]{EAF8E0}Activity2 & \cellcolor[HTML]{EAF8E0}Activity3 & \cellcolor[HTML]{EAF8E0}Activity4 & \cellcolor[HTML]{EAF8E0}Activity5 & \cellcolor[HTML]{EAF8E0}Activity6 & \cellcolor[HTML]{EAF8E0}Activity7 & \cellcolor[HTML]{EAF8E0}Activity8 & \cellcolor[HTML]{EAF8E0}Activity9 & \cellcolor[HTML]{EAF8E0}Activity10 \\ \hline
Activity1                  & \cellcolor[HTML]{FFFFC7}95.45     & 0                                 & 0                                 & 0                                 & 0                                 & 0                                 & 0                                 & 0                                 & 0                                 & 4.54                               \\ \hline
Activity2                  & 0                                 & \cellcolor[HTML]{FFFFC7}84        & 0                                 & 0                                 & 0                                 & 0                                 & 0                                 & 4                                 & 0                                 & 12                                 \\ \hline
Activity3                  & 0                                 & 0                                 & \cellcolor[HTML]{FFFFC7}100       & 0                                 & 0                                 & 0                                 & 0                                 & 0                                 & 0                                 & 0                                  \\ \hline
Activity4                  & 0                                 & 0                                 & 0                                 & \cellcolor[HTML]{FFFFC7}91.67     & 0                                 & 0                                 & 0                                 & 0                                 & 8.33                              & 0                                  \\ \hline
Activity5                  & 0                                 & 0                                 & 0                                 & 0                                 & \cellcolor[HTML]{FFFFC7}100       & 0                                 & 0                                 & 0                                 & 0                                 & 0                                  \\ \hline
Activity6                  & 0                                 & 0                                 & 0                                 & 0                                 & 0                                 & \cellcolor[HTML]{FFFFC7}89.65     & 10.35                             & 0                                 & 0                                 & 0                                  \\ \hline
Activity7                  & 0                                 & 0                                 & 0                                 & 0                                 & 0                                 & 0                                 & \cellcolor[HTML]{FFFFC7}100       & 0                                 & 0                                 & 0                                  \\ \hline
Activity8                  & 0                                 & 4                                 & 0                                 & 0                                 & 0                                 & 0                                 & 0                                 & \cellcolor[HTML]{FFFFC7}92        & 0                                 & 4                                  \\ \hline
Activity9                  & 0                                 & 0                                 & 0                                 & 0                                 & 0                                 & 0                                 & 0                                 & 0                                 & \cellcolor[HTML]{FFFFC7}100       & 0                                  \\ \hline
Activity10                 & 0                                 & 4                                 & 0                                 & 0                                 & 0                                 & 0                                 & 0                                 & 24                                & 0                                 & \cellcolor[HTML]{FFFFC7}72         \\ \hline
\end{tabular}
\end{table*}
The confusion matrix shows that Activity 3, 5, 7, and 9 remain the best recognized classes amongst all the classes considered in the study. The classifier is mostly confused between activities 8 and 10, resulting in poor classification accuracies for these two cases. This is likely because these signatures share common features in the micro-Doppler latent feature space due to the proximity between their motion characteristics. In addition, both these signatures possess alternating micro-Doppler information, which confuses the network. The overall classification accuracy attained for this case is 92.5\%.

Another possible reason for low classification performance for this case could be the volume of data used to train the classifier. We use 300 measurements (30 measurements for each activity) for training and the remaining 300 for testing, which we believe is insufficient for training the network. Therefore, in the next section, we present the classification results for cases where the measurement data is augmented with the simulated data.  

\subsection{Data Augmentation Scenario}
\begin{figure*}[htbp]
\centering
\includegraphics[scale=0.75]{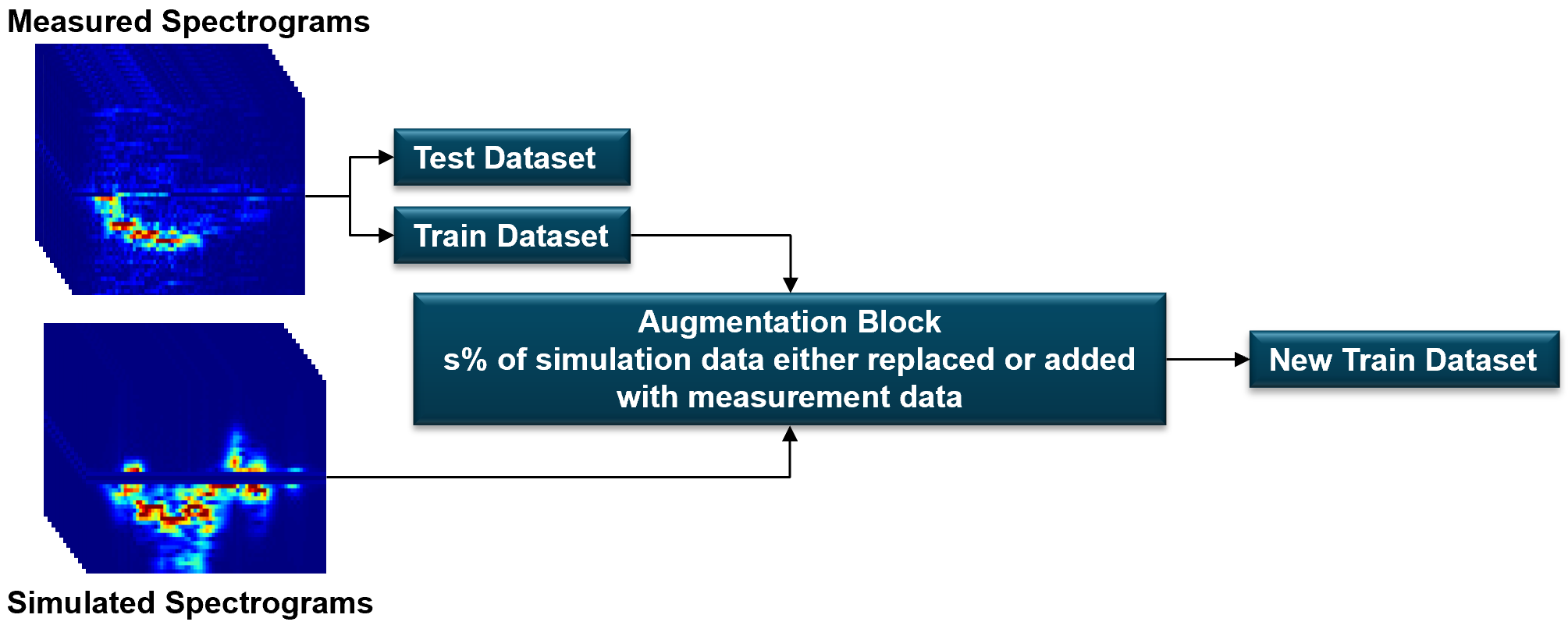}
\caption{Augmentation study: The ratio of simulation data either replaced/augmented with measurement data is varied to study the impact of data replacement/augmentation on the classification performance. s=0,  corresponds to case 1 where both training and test data comprise of measured spectrograms only (TMTM).}
\label{fig:augstudy}
\end{figure*}

In this section, we test two data augmentation schemes, replacement and the augmentation as shown in Fig. \ref{fig:augstudy}. In the replacement scheme, we replace a part of measurement data with simulation data, keeping the overall dataset size of the training data the same.  The purpose of investigating this scheme is to see whether the performance can be improved by replacing the unbalanced or low-quality measurement data with good simulation data.
On the other hand, unlike replacement, the augmentation scheme adds additional simulation data to the measurement dataset, \textit{increasing the training dataset's overall size}. 

\subsubsection{Case 2: Style Transfer Based Data Augmentation}
This section investigates two data augmentation scenarios: first, 60\% of the measurement data is replaced with simulated spectrograms generated through the style transfer framework. Second, 60\% of the style transferred dataset is added to the measurement data, resulting in increased overall training dataset size. We report our replacement scheme results in Table \ref{table:repstyle} and the augmentation scheme results in Table \ref{table:augstyle}.

We achieve an overall classification accuracy of 96.2\% in the replacement scenario and 97.3\% in the replacement scenario, nearly 3.5\% and 4.6\% greater than the TMTM case (that is, when only measurement data is used for both training and testing) in the two augmentation scenarios considered. Furthermore, the confusion matrices clearly show that activities 8 and 10 are now more discernible and have improved classification accuracies compared to the TMTM case. One plausible reason for the improvement in the replacement scheme could be substituting some of the noisy measurement samples with the excellent style transferred data that possess good motion characteristics and environmental effects. On the other hand, the improvement in the data augmentation scheme could be attributed to the fact that the measurement training support size is increased, giving the neural network enough data to be trained well and extract more features from this diverse dataset. 
\begin{table*}[]
\centering
\caption{Confusion matrix for replacement scenario where 60\% measurement dataset is replaced with style transferred dataset for training and remaining 50\% only measurement is used for testing the performance of the network.}
\label{table:repstyle}
\begin{tabular}{|
>{\columncolor[HTML]{EAF8E0}}c |c|c|c|c|c|c|c|c|c|c|}
\hline
True Class/Predicted Class & \cellcolor[HTML]{EAF8E0}Activity1 & \cellcolor[HTML]{EAF8E0}Activity2 & \cellcolor[HTML]{EAF8E0}Activity3 & \cellcolor[HTML]{EAF8E0}Activity4 & \cellcolor[HTML]{EAF8E0}Activity5 & \cellcolor[HTML]{EAF8E0}Activity6 & \cellcolor[HTML]{EAF8E0}Activity7 & \cellcolor[HTML]{EAF8E0}Activity8 & \cellcolor[HTML]{EAF8E0}Activity9 & \cellcolor[HTML]{EAF8E0}Activity10 \\ \hline
Activity1                  & \cellcolor[HTML]{FFFFC7}100       & 0                                 & 0                                 & 0                                 & 0                                 & 0                                 & 0                                 & 0                                 & 0                                 & 0                                  \\ \hline
Activity2                  & 0                                 & \cellcolor[HTML]{FFFFC7}88        & 0                                 & 0                                 & 0                                 & 0                                 & 0                                 & 12                                & 0                                 & 0                                  \\ \hline
Activity3                  & 0                                 & 0                                 & \cellcolor[HTML]{FFFFC7}100       & 0                                 & 0                                 & 0                                 & 0                                 & 0                                 & 0                                 & 0                                  \\ \hline
Activity4                  & 0                                 & 0                                 & 0                                 & \cellcolor[HTML]{FFFFC7}100       & 0                                 & 0                                 & 0                                 & 0                                 & 0                                 & 0                                  \\ \hline
Activity5                  & 0                                 & 0                                 & 0                                 & 0                                 & \cellcolor[HTML]{FFFFC7}100       & 0                                 & 0                                 & 0                                 & 0                                 & 0                                  \\ \hline
Activity6                  & 6.9                               & 0                                 & 0                                 & 0                                 & 0                                 & \cellcolor[HTML]{FFFFC7}89.7      & 3.45                              & 0                                 & 0                                 & 0                                  \\ \hline
Activity7                  & 0                                 & 0                                 & 0                                 & 0                                 & 0                                 & 0                                 & \cellcolor[HTML]{FFFFC7}100       & 0                                 & 0                                 & 0                                  \\ \hline
Activity8                  & 0                                 & 0                                 & 0                                 & 0                                 & 0                                 & 0                                 & 0                                 & \cellcolor[HTML]{FFFFC7}96        & 0                                 & 4                                  \\ \hline
Activity9                  & 0                                 & 0                                 & 0                                 & 5.7                               & 0                                 & 2.8                               & 0                                 & 0                                 & \cellcolor[HTML]{FFFFC7}91.5      & 0                                  \\ \hline
Activity10                 & 0                                 & 0                                 & 0                                 & 0                                 & 0                                 & 0                                 & 0                                 & 8                                 & 0                                 & \cellcolor[HTML]{FFFFC7}92         \\ \hline
\end{tabular}
\end{table*}
\begin{table*}[]
\centering
\caption{Confusion matrix for augmentation scenario where 60\% style transferred dataset is added to the training measurement dataset. The remaining 50\% measurement is used for testing the performance of the network.}
\label{table:augstyle}
\begin{tabular}{|
>{\columncolor[HTML]{EAF8E0}}c |c|c|c|c|c|c|c|c|c|c|}
\hline
True Class/Predicted Class & \cellcolor[HTML]{EAF8E0}Activity1 & \cellcolor[HTML]{EAF8E0}Activity2 & \cellcolor[HTML]{EAF8E0}Activity3 & \cellcolor[HTML]{EAF8E0}Activity4 & \cellcolor[HTML]{EAF8E0}Activity5 & \cellcolor[HTML]{EAF8E0}Activity6 & \cellcolor[HTML]{EAF8E0}Activity7 & \cellcolor[HTML]{EAF8E0}Activity8 & \cellcolor[HTML]{EAF8E0}Activity9 & \cellcolor[HTML]{EAF8E0}Activity10 \\ \hline
Activity1                  & \cellcolor[HTML]{FFFFC7}100       & 0                                 & 0                                 & 0                                 & 0                                 & 0                                 & 0                                 & 0                                 & 0                                 & 0                                  \\ \hline
Activity2                  & 0                                 & \cellcolor[HTML]{FFFFC7}84        & 0                                 & 0                                 & 0                                 & 0                                 & 0                                 & 8                                 & 0                                 & 8                                  \\ \hline
Activity3                  & 0                                 & 0                                 & \cellcolor[HTML]{FFFFC7}100       & 0                                 & 0                                 & 0                                 & 0                                 & 0                                 & 0                                 & 0                                  \\ \hline
Activity4                  & 0                                 & 0                                 & 0                                 & \cellcolor[HTML]{FFFFC7}100       & 0                                 & 0                                 & 0                                 & 0                                 & 0                                 & 0                                  \\ \hline
Activity5                  & 0                                 & 0                                 & 0                                 & 0                                 & \cellcolor[HTML]{FFFFC7}100       & 0                                 & 0                                 & 0                                 & 0                                 & 0                                  \\ \hline
Activity6                  & 0                                 & 0                                 & 0                                 & 0                                 & 0                                 & \cellcolor[HTML]{FFFFC7}100       & 0                                 & 0                                 & 0                                 & 0                                  \\ \hline
Activity7                  & 0                                 & 0                                 & 0                                 & 0                                 & 0                                 & 0                                 & \cellcolor[HTML]{FFFFC7}100       & 0                                 & 0                                 & 0                                  \\ \hline
Activity8                  & 0                                 & 0                                 & 0                                 & 0                                 & 0                                 & 0                                 & 0                                 & \cellcolor[HTML]{FFFFC7}92        & 0                                 & 8                                  \\ \hline
Activity9                  & 0                                 & 5.7                               & 0                                 & 0                                 & 0                                 & 0                                 & 0                                 & 0                                 & \cellcolor[HTML]{FFFFC7}94.3      & 0                                  \\ \hline
Activity10                 & 0                                 & 0                                 & 0                                 & 0                                 & 0                                 & 0                                 & 0                                 & 4                                 & 0                                 & \cellcolor[HTML]{FFFFC7}96         \\ \hline
\end{tabular}
\end{table*}
The results obtained for both studies demonstrate that the quality of signatures generated through the style transfer framework possesses excellent kinematic fidelity with the measurement data and can capture the various environmental factors. In the next section, we benchmark the data replacement and augmentation performances across four synthesized datasets: no noise simulated dataset, AWGN noise dataset, GAN noise dataset, and style transferred dataset. 

\subsection{Performance Benchmarking Across Multiple Scenarios}
\begin{figure}[t]
\centering
\includegraphics[scale=0.5]{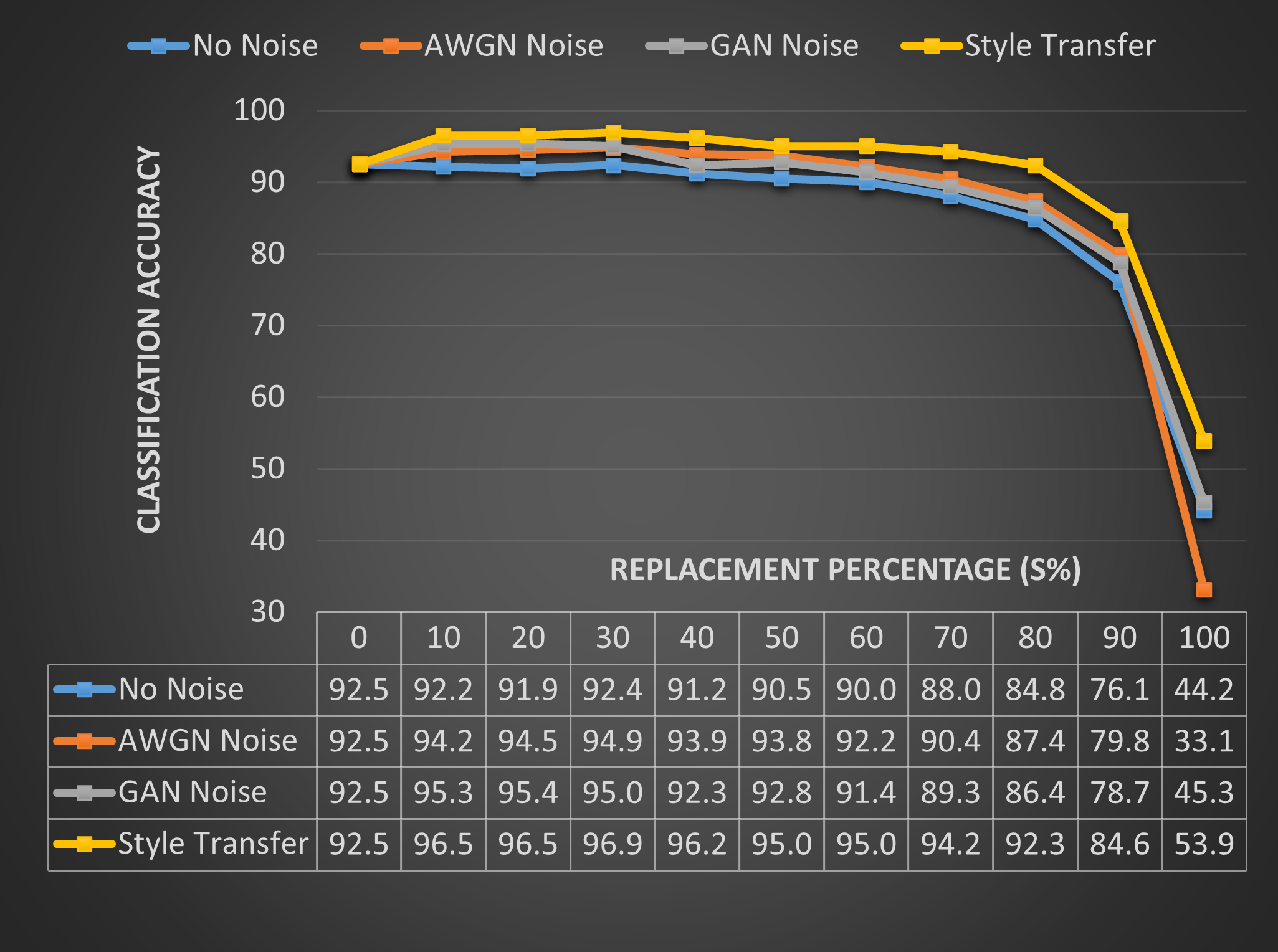}
\caption{Replacement Results: Classification accuracies as a function of the percentage of replacement data for four datasets-clean spectrograms, AWGN added spectrograms, GAN noise added spectrograms, and style transferred spectrograms. Note that the training dataset remains the same.}
\label{fig:replacementfig}
\end{figure}

\begin{figure}[t]
\centering
\includegraphics[scale=0.5]{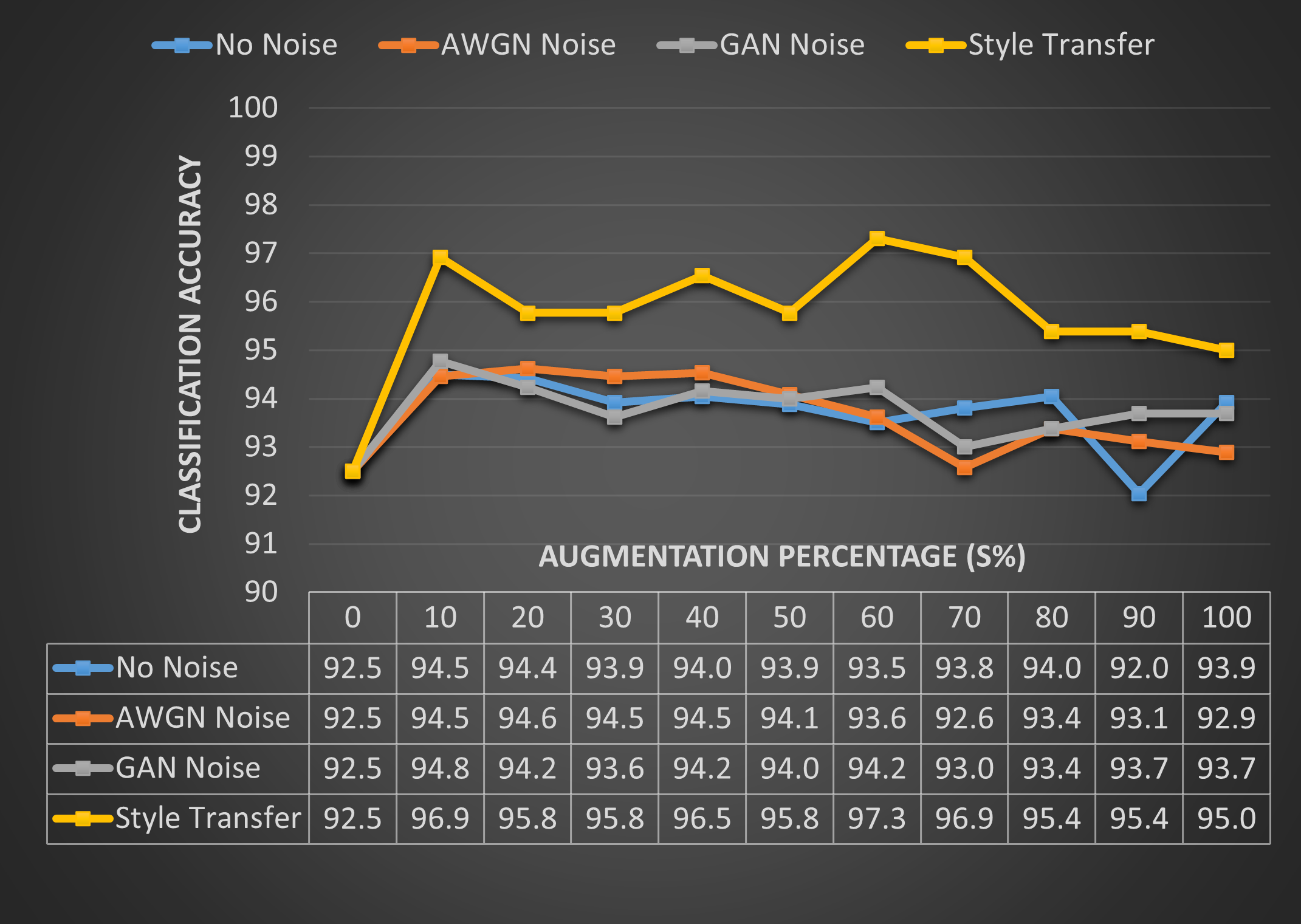}
\caption{Augmentation Results: Classification accuracies as a function of the percentage of augmented data for four datasets- no noise clean spectrograms, AWGN added spectrograms, GAN noise added spectrograms and style transferred spectrograms. Note that the training dataset changes with the percentage of simulated data (s) augmented with the measurement dataset.}
\label{fig:augmentationfig}
\end{figure}

We benchmark the proposed style transfer-based data augmentation performance with three synthetic datasets: the noise simulated dataset, the AWGN noise dataset, and the GAN noise dataset. For this, we gradually increase $s$, simulation data percentage from $0\%$ to $100\%$, for both replacement and augmentation studies in each data augmentation scenario. We present the benchmarking results for replacement in Fig. \ref{fig:replacementfig} and augmentation in Fig. \ref{fig:augmentationfig}. We observe from the results that the style transferred simulation data-based augmentation has the highest performance across all the $s$ (the percentage of simulation data). In addition, the performance is better than the TMTM case. Note that $s=0$ indicates the TMTM case.  The performance for the no-noise case is the worst; that is when we use clean spectrograms for the data augmentation. The classification performances for AWGN and GAN are comparable. For some cases, the performance of AWGN added data is better than GAN noise-based data, and for others, GAN-based data offers better performance. The classification results are in perfect agreement with our qualitative and quantitative analysis of the different datasets.

As can be seen from Fig. \ref{fig:replacementfig}, the classification accuracies decrease with an increase in $s$. This indicates the case when measurement data is replaced with more and more simulation data. Interestingly, the performance for style transfer data remains equivalent to the TMTM case even when 80\% of the measurement data is replaced with this synthetic data indicating the quality of signatures matching the measurement data. The classification accuracy is nearly 8\% greater than other cases. We can draw similar inferences from the augmentation results presented in Fig.\ref{fig:augmentationfig}. When more and more simulation data is added, the performance improves for almost all the cases; however, it is more prominent in style transferred data. 

Note that the proposed style transfer framework generates the synthetic signatures to be used for providing training support to the limited data and not for testing the classification performance. Therefore, it does not affect the real-time classification performance of the ISAC system. 
\section{Conclusion}
\label{sec:conclusion}
This paper presents an effective style transfer framework to synthesize realistic micro-Doppler signatures that possess excellent motion characteristics and environmental factors such as noise, multipath, clutter, as well as the target-dependent nuances such as occlusion effects. The proposed network extracts global motion content from clean simulated signatures and background texture information directly from measured signatures to form a third image termed as \textit{style transferred image}, possessing the qualities of two signatures. To further demonstrate the quality of synthesized signatures, we performed a detailed qualitative and quantitative analysis by visual inspection of spectrograms and its latent feature space relative to the original measurement dataset. 

We also benchmark the approach with three other synthesized datasets: clean simulation data with no noise, AWGN noise added dataset and GAN noise dataset. The results highlight the superior quality of the style transferred signatures. Additionally, we propose a novel data augmentation scheme, which is a potential application of these signatures. We test the classification performance under the following augmentation scenarios: measurement data augmented with simulation data, with no noise, simulation data with added AWGN, simulation data with GAN noise, and style transferred data. The results show that the data generated through the style transfer framework outperformed all other cases by 3-5\% on average. The improvement is more pronounced ($\geq 8\%$), especially when the replacement percentage is more than 80\%. 

Overall, the paper demonstrates the feasibility of generating realistic simulated micro-Doppler spectrograms using a style transfer framework. Since these signatures can effectively mimic realistic signatures, they can be used to augment the training dataset and effectively enhance the sensing performance of existing ISAC system for real-world applications such as e-healthcare and ambient assisted living without degrading its communication capabilities. The main idea is to use the existing communication platforms to implement an augmented sensing system (with minimal or no modifications to the existing communication system). 

\noindent\textbf{Future Directions:}\\
Our current network obtains the stylized image through multiple forward pass and error back-propagation through the network for each image. In the future, we plan to use faster neural style transfer to generate the synthetic dataset using a single forward pass through the network. Furthermore, we will investigate the generalization capability of the proposed style transfer network and the proposed augmentation scheme in different environmental conditions such as different rooms, different IEEE 802.11 standards and include measurements from through-the-wall scenarios. We believe the style transfer technique can be easily scaled to other signatures types and opens up opportunities in understanding the natural phenomena directly from the measurement data. Therefore, we plan to extend our research to using a channel state information (CSI) based measurements from the ISAC system to generate the micro-Doppler signatures and the corresponding simulated signatures using CSI-based SimHumalator (currently under development). Interested researchers can download the passive ISAC based simulator from \url{https://uwsl.co.uk/} and can get the latest updates on the development of our CSI-based SimHumalator from the same website.  
\section*{Acknowledgments}
This work is part of the OPERA project funded by the UK Engineering and Physical Sciences Research Council (EPSRC), Grant No: EP/R018677/1. 
\bibliographystyle{IEEEtran}
\bibliography{References}
\vspace{-1cm}
\end{document}